\documentstyle[aps,pre,multicol,epsf]{revtex}

\begin{document}

\draft

\title{Anomalous Scaling of Fracture Surfaces}
\author{Juan M. L\'opez$^1$ and Jean Schmittbuhl$^2$}

\address{$^1$ Department of Mathematics, Imperial College, 
180 Queen's Gate, 
London SW7 2BZ,
United Kingdom}

\address{$^2$ Lab. de G\'eologie, URA 1316, Ecole Normale Sup\'erieure,
24 rue Lhomond, 75231 Paris Cedex 05, France}
\maketitle

\begin{abstract}
  We argue that fracture surfaces may exhibit {\em anomalous} dynamic
  scaling properties akin to what occurs in some models of kinetic
  roughening. We determine the complete scaling behavior of the local
  fluctuations of a brittle fracture in a
  granite block from experimental data. 
  We obtain a global roughness exponent $\chi = 1.2$ which
  differs from the local one, $\chi_{loc} = 0.79$. Implications on
  fracture physics are discussed.

\end{abstract}


\begin{multicols}{2}
\narrowtext
  
  The study of the morphology of fracture surfaces is nowadays a very
  active field of research. From the early work of Mandelbrot {et. al.}
  \cite{mandelbrot}, much effort has been paid to the statistical
  characterization of the resulting fractal surfaces in fracture processes.
  Scale invariance has been found in many experiments
  \cite{mandelbrot,fractalD,maloy,poon,jean,kertesz1,poirier,kertesz2,engoy}
  and it is now well established that, in general, crack surfaces exhibit
  self-affine scaling properties in a large range of length scales (see
  \cite{meakin} for a more detailed account of experiments).

A self-affine surface $h(\hbox{\bf x})$ is invariant under an anisotropic
scale transformation, in the sense that $h(\hbox{\bf x})$ has the same
statistical properties as $\lambda ^{-\chi} h(\lambda \hbox{\bf x})$, where
$\chi$ is the {\em roughness exponent}.  Initially, the hope in studying
the surface morphology was to relate geometry to mechanical properties
(toughness, plasticity, etc) in order to obtain a material characterization
by means of the roughness exponent. However, experimental results in very
different types of materials (from ductile aluminium alloys to brittle
materials like rock) seem to support the idea of a {\em universal}
roughness exponent which is very independent of the material properties.
For three-dimensional fractures, an exponent $\chi (3D) \simeq 0.8-0.9$
\cite{mandelbrot,fractalD,maloy,poon,jean} has been measured, whereas in
dimension two $\chi (2D) \simeq 0.6-0.7$
\cite{kertesz1,poirier,kertesz2,engoy}.  It seems reasonable to expect that
material properties should affect the fracture surface roughness. In
particular, toughness and anisotropy should be relevant for the fracture
crack morphology.  However, the above mentioned experimental results seem
to lead to the surprising conclusion that there is no correlation of $\chi$
with mechanical properties.

The treatment of the fracture crack as a self-affine rough surface leads in
a natural way to the close field of kinetic roughening (see
\cite{rev-rough} for recent reviews in the subject). A direct
mapping of the crack at the stationary state into the Kardar-Parisi-Zhang
equation \cite{kpz} has been proposed in Ref. \cite{universal}.  The crack
surface has also been considered as the trace of the crack front whose
propagation is modeled by other types of non-linear Langevin equations
\cite{blp-prl}.

In this Letter, we will show that crack surfaces have very much in common
with those obtained in growth processes exhibiting what is called an {\em
  anomalous} dynamic scaling.  We argue that the scaling of the
fluctuations of crack surfaces is, in the sense of kinetic roughening, {\em
  intrinsically anomalous} rather than a simple Family-Vicsek one. 
The analysis of the roughness data of a crack experiment demonstrates the
validity of our argument. Physical consequences of 
this scaling for fracture will be discussed. 

{\em Family-Vicsek scaling.--} In practice, the self-affine character of a
surface in dimension $d+1$ is shown by studying the scaling of the
fluctuations of the surface height over the whole system of total size $L$.
The invariance property under the scale transformation implies that the
{\em global} width at time $t$, 
$W(L,t) = \langle \bigl(h(\hbox{\bf x},t) -
\overline h(t) \bigr)^2 \rangle_{\hbox{\bf x}} ^{1/2}$ where 
$\hbox{\bf x}$ is the position,
has to scale as \cite{fv}
\begin{equation}
\label{FV-globalwidth}
W(L,t) = L^{\chi} f(L/t^{1/z}),
\end{equation}
where $z$ is the dynamical exponent and $f(u)$ is the scaling function 
\begin{equation}
\label{FV-forf}
f(u) \sim
\left\{ \begin{array}{lcl}
    \hbox{\rm const.} & \hbox{\rm if} & u \ll 1\\
    u^{-\chi} & \hbox{\rm if} &  u \gg 1.
\end{array}
\right.
\end{equation}
$\chi$ is the roughness exponent and gives the scaling of the surface in
saturation, $W(L,t \gg L^z) \sim L^\chi$.  These two exponents characterize
the universality class of the particular growth model.  Equivalently, the
scaling behavior of the surface might be obtained by looking at the {\it
  local} width over a window of size $l \ll L$:
\begin{equation}
\label{local-FV}
w(l,t) \sim  
\left\{ \begin{array}{lcl}
     t^{\chi/z} & \hbox{\rm if} &  t \ll l^z\\
     l^{\chi} & \hbox{\rm if} &  t \gg l^z.
     \end{array}
\right.
\end{equation} 
This is the method actually 
used in real experiments where the size of the system
$L$ remains constant and the fluctuations are calculated over scales $l \ll
L$. Note that the local width saturates at time $l^z$ as $w(l,t \gg
l^z) \sim l^{\chi}$ independently of the system size.

A complementary technique to determine the critical exponents of a growing
surface is to study the Fourier transform of the interface height in a
system of linear size $L$, $\widehat{h}(\hbox{\bf k},t) = L^{-d/2}
\sum_{\hbox{\bf x}} [h(\hbox{\bf x},t) - \overline h(t)] \exp(i\hbox{\bf
  k}\hbox{\bf x})$, where the spatial average of the height has been
subtracted.  In this representation, the properties of the surface can be
investigated by calculating the power spectrum $S(k,t) = \langle
\widehat{h}(\hbox{\bf k},t) \widehat{h}(-\hbox{\bf k},t) \rangle $, which
contains the same information on the system as the local width. In most
growth models, the power spectrum scales as
\begin{equation}
\label{Sscal}
S(k,t) =k^{-(2\chi+d)} s(kt^{1/z}),
\end{equation}
where $s$ is the simple scaling function 
\begin{equation}
\label{FVforS}
s(u) \sim  
\left\{ \begin{array}{lcl}
     \hbox{\rm const.} & \hbox{\rm if} & u \gg 1\\
     u^{2\chi + d} & \hbox{\rm if} &  u \ll 1.
     \end{array}
\right. 
\end{equation}
This form for the power spectrum can be easily inverted to obtain the
scaling behavior of both local and global widths described above.

{\em Anomalous scaling and implications.--} In very recent studies of
several growth models \cite{anom-ref,lack}, it has been found that the
surface fluctuations may exhibit an {\em anomalous} scaling, in the sense
that, although the global width behaves as in
(\ref{FV-globalwidth})-(\ref{FV-forf}) local surface fluctuations do not
satisfy Eq.(\ref{local-FV}), but scale as
\begin{equation}
w(l,t) \sim  
\left\{ \begin{array}{lcl}
     t^{\chi/z} & \hbox{\rm if} & t \ll l^z\\
     t^{\beta_*} \; l^{\chi_{loc}} & \hbox{\rm if} & t \gg l^z,
     \end{array}
\right.
\label{local-anom}
\end{equation} 
where the exponent $\beta_*=(\chi - \chi_{loc})/z$ is an anomalous time
exponent and $\chi_{loc}$ the local roughness exponent.  Thus, in the case
of anomalous scaling two exponents, $\chi_{loc}$ and $\chi$, enter the
scaling and must be taken into account to give a complete description of
the scaling behavior of the surface. An outstanding consequence is that the
local width does not saturate at times $l^z$ but when the whole system
does, {\em i.e.} at times $L^z$, giving an unconventional dependence of the
stationary local width on the system size as
\begin{equation}
\label{saturated-loc}
w(l,t \gg L^z) \sim l^{\chi_{loc}} L^{\chi-\chi_{loc}},
\end{equation}
in such a way that the magnitude of the roughness over regions of same size
$l$ at saturation is not just a function of the window size but also of the
system size, which is distinctly different from what happens in the
Family-Vicsek case.
 
Owing to several experimental limitations, anomalous scaling is
difficult to observe (see \cite{exp} for kinetic roughening experiments
in which anomalous scaling was found). 
Since the system size $L$ of experiments can hardly
be changed over a broad range, the dependence of the global width
on the system size cannot be actually determined.
Only local fluctuations over a window $l$ can
be measured.  Moreover, very often the time evolution of a crack cannot be
monitored and in most experiments only the final crack surface is 
analyzed, {\em i.e.} Eq.(\ref{saturated-loc}) for a fixed system size
$L$. This immediately leads to the conclusion that, whether anomalous
scaling exists, only $\chi_{loc}$ is actually at reach of the methods
currently used in experiments.  

In terms of the power spectrum, the existence of a local exponent
$\chi_{loc} \ne \chi$ comes from a nonstandard form of the scaling function
$s(u)$ in Eq.(\ref{Sscal}). It has recently been shown \cite{intrinsic}
that the anomalous scaling (\ref{local-anom}) is associated with either
{\em super-roughening} \cite{super-rough} or a power spectrum that
satisfies the dynamic scaling behavior stated in (\ref{Sscal}) but with a
distinct scaling function
\begin{equation}
\label{Anom-S}
s(u) \sim  
\left\{ \begin{array}{clc}
     u^{2(\chi-\chi_{loc})} & \hbox{\rm if} & u \gg 1\\
     u^{2\chi + d} & \hbox{\rm if} & u \ll 1.
     \end{array}
\right. 
\end{equation}
So, in the stationary regime (at times $t \gg L^z$) the power spectrum
scales as $S(k,t) \sim k^{-(2\chi_{loc} + d)}
L^{2(\chi-\chi_{loc})}$, and not simply as $k^{-(2\chi + d)}$ as
corresponds to a standard scaling.  This means that experimental
determinations of the roughness exponent of fracture crack surfaces from
the decay of the power spectrum with $k$ also give a measure of
$\chi_{loc}$ and not $\chi$. 

Most of the experimental studies are unable to follow the crack in time and
much important information about the complete scaling is lost.  In the
majority of the experiments one has to deal with a static fracture surface
and fluctuations of its height are evaluated over windows of different
sizes $l$.  So, {\em neither} (\ref{local-FV}) {\em nor} (\ref{local-anom})
{\em scaling forms are actually tested in fracture experiments}.

{\em Experiment.--} 
In the following we present an analysis of the data describing growth
of the crack roughness.  In this experiment a fracture was initiated from a
straight notch in a granite sample ($25$cm$\times 25$cm$\times
12$cm)\cite{jean}. It is a mode I unstable crack. The crack roughness
increases from two hundredth of a millimeter to several millimeters.
Topographies of two areas of $5$cm$\times 4$cm were recorded with a first
mechanical profiler along a regular grid (100 parallel profiles). The $x$
direction which is parallel to the initial notch was sampled with $1050$
points ($\Delta x = 50 \mu$m). The grid step along the perpendicular
direction (i.e. the crack propagation direction) was $\Delta y = 350\mu$m.
A third map ($5$cm$\times 5$cm) was obtained from a second and independent
mechanical profiler with a higher resolution.  Two hundred parallel
profiles were recorded with $2050$ points per profile (in which  
$\Delta x =32.5 \mu$m and $\Delta y = 250\mu$m). 
We assumed that the crack speed was constant which translates in a linear
relationship between position $y$ and time $t$. As a consequence, we
consider the one-dimensional profiles as descriptions of the advancing
crack $h(x,t)$.  The complete spatio-temporal behavior of the surface can
thus be obtained. 
 
In reference\cite{jean}, the scaling form (\ref{local-FV}) was checked for
the two first data sets.  However, a careful inspection of the data
collapse reported, Fig. 1 in \cite{jean}, reveals that the scaling function
goes like a power law for large abscissa values rather than be a constant.
Also the slope for small values of the argument does not match well with
the correct value.  As we will see much better and more accurate results
are obtained if, instead of assuming a Family-Vicsek behavior, we analyze
the data on the basis of an anomalous scaling. In this case, from
(\ref{local-anom}) it is easy to see that the corresponding scaling
function would be
\begin{equation}
\label{f-anom}
g_A(u) \sim 
\left\{ \begin{array}{clc}
     u^{-(\chi-\chi_{loc})} & \hbox{\rm if} & u \ll 1\\
     u^{-\chi} & \hbox{\rm if} & u \gg 1.
     \end{array}
\right., 
\end{equation}
in such a way that $w(l,t)/l^\chi = g_A(l/t^{1/z})$,
where the label $A$ denotes the anomalous scaling form. 

In Figure 1 we present the data collapse of $w(l,t)/l^\chi$ {\it vs.}
$l/t^{1/z}$ for the high resolution map of the crack surface obtained in
the experiment. The best data collapse occurs for a global roughness
exponent $\chi=1.2\pm 0.1$ and $z=1.2\pm 0.15$.  The same results were also
found when the other two lower resolution data sets were used. 
The non-constant behavior for $u \ll 1$ in Fig. 1 
is the main fingerprint of the
anomalous character of the scaling. Figure 1 is in excellent agreement with
a scaling function like (\ref{f-anom}).

According to (\ref{f-anom}) the power law $u^{-0.41}$ for $u \ll 1$ in Fig.
1 corresponds to a local roughness exponent $\chi_{loc}=0.79$. This value
can be compared with the value obtained from the long time behavior of the
height difference correlation function, $G(l,t) = \langle \bigl( h(x+l,t) -
h(x,t) \bigr)^2 \rangle_x^{1/2}$, that we plot in Figure 2 for the long
times limit, $t \gg l^z$, and the highest resolution data.  At long times,
$G(l)$ displays a power law behavior $l^{\chi_{loc}}$ that gives an
independent determination of the {\em local} roughness exponent.  As shown
in Fig. 2, data fit to $\chi_{loc} = 0.79$ in agreement with the value
obtained from the slope $-0.41$ in Fig. 1.  This estimate of $\chi_{loc}$
was confirmed by several other independent techniques: variable band width,
return probability and wavelet analysis \cite{sa_tec,sa_wav}.

We have also calculated the power spectrum and in Figure 3 we plot $S(k,t)$
{\it vs.} $k$ in a log-log plot for times $t=50$,75 and 100 for the high
resolution data. The curves are clearly shifted in time as corresponds to a
power spectrum scaling function like Eq.(\ref{Anom-S}) and not the
Family-Vicsek one in (\ref{Sscal}).  $S(k,t)$ decays with a power law
$k^{-2.58}$ which is consistent with $k^{-(2\chi_{loc} + 1)}$ and
$\chi_{loc} = 0.79$. We were unable to obtain a collapse of good quality
form these data.

{\em Conclusions.-} In this study we have shown that an anomalous dynamic
scaling, (\ref{local-anom}) or (\ref{saturated-loc}), captures much better
the features of the crack geometry than the standard Family-Vicsek one.  We
obtained for an unstable brittle fracture of a granite block a global
exponent $\chi = 1.2$, which is different from the local exponent
$\chi_{loc}=0.79$.  The very robust value of the local exponent has been
demonstrated in the past.

The existence of two different and independent roughness exponents that
characterize the complete scaling of a surface, as it follows from
Eq.(\ref{saturated-loc}), has important implications in the appearance (and
geometry) of anomalously roughened surfaces. To illustrate this point we
plot in Figure 4 two surfaces with exactly the same local , $\chi_{loc} =
1/2$, but a different global exponent, $\chi = 3/4$ and $1/2$. The two
interfaces plotted in Fig. 4 are quite different to the naked eye despite
they have the same local roughness exponent. From Figure 4 it is clear that
$\chi_{loc}$ necessarily gives just a part of the information about the
scaling when anomalous roughening exists and indeed the global exponent
$\chi$ does give account of the large peaks taken by the surface in the
anomalous case ({\it i.e.} when $\chi \ne \chi_{loc}$).  The similarity
with what happens to the patterns found in experimental cracks (see for
instance Fig 1. in \cite{engoy}) suggests that they may exhibit the same
type of double scaling.  More precisely, Fig. 1 in Ref.\cite{engoy} shows
that the resulting surfaces in wood for tangential and radial fractures
display strong different morphologies with the same local exponent
$\chi_{loc} \simeq 0.68$.  These authors found {\em puzzling} the large
peaks taken by tangential fracture surfaces in clear contrast to the quite
flat look of radial fractures, very much as occurs in the example we plot
in Figure 4.  Also Zhang {\em et. al.} \cite{anis-model} in a numerical
model of fracture in anisotropic materials found a local exponent roughly
constant $\chi_{loc} \simeq 0.7$ {\em surprisingly} independent of the
orientation of the fracture, although visible differences in the surfaces
were noted (see Fig. 2 in Ref. \cite{anis-model}). Experiments in material
showing different rupture modes by Bouchaud {\em et. al.} \cite{fractalD}
gave similar results. We believe that a determination of the global
roughness exponent in all these experiments could allow a better
characterization of the fracture surface morphology and its relationship
with material properties.

A second physical consequence of the anomalous scaling is that the
saturation roughness is not only function of window size but also of the
system size. Implications on the crack process are important. Information
about the system size exists along the crack front during the propagation
even at Rayleigh speed. This may illustrate the role of interactions
between elastic waves and front propagation for the geometry of the front.

More studies should be initiated to check weather or not the global
roughness exponent is in general a valid index to characterize fracture
surfaces.

JML wishes to thank J.\ V.\ Andersen for discussions and encouragement and
R.\ Cuerno for a careful reading of the manuscript.  JS thanks K.J. M\aa
l\o y for very fruitful discussions.  This work has been supported by the MEC
of the Spanish Government, the European Commission, the Ecodev-CNRS program
and the GDR ``G\'eom\'ecanique des Roches Profondes''.

\begin{figure}
\centerline{
\epsfxsize=5cm
\epsfbox{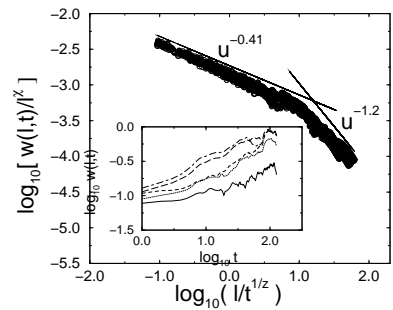}}
\caption{
  Inset shows the local width {\it vs.} time for the higher resolution data
  of the crack experiment calculated over windows of sizes $l=100$ (solid),
  300 (dotted), 500 (dashed), 700 (long-dashed) and 900 (dot-dashed) 
  (all in units of the grid step $\Delta x$).  
  Data are collapsed for $\chi = 1.2$ and
  $z = 1.2$ in the main panel for windows sizes ranging from $l=10$ to
  $l=1200$.  The non constant behavior at small values of the argument
  (main panel) reflects the anomalous character of the scaling, which
  agrees with Eq.(9). }
\end{figure}

\begin{figure}
\centerline{
\epsfxsize=5cm
\epsfbox{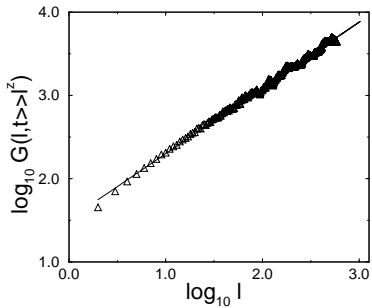}}
\caption{Height difference correlation function for the experimental
data displayed in Fig. 1. 
The straight line is a fit of the data and its slope 0.79 gives a
determination of the local roughness exponent.}
\end{figure}

\begin{figure}
\centerline{
\epsfxsize=5cm
\epsfbox{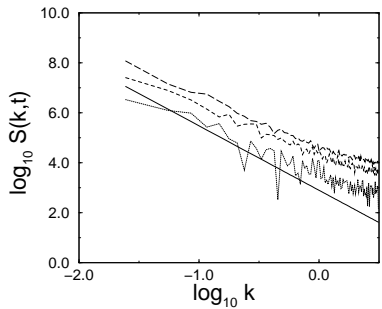}}
\caption{Power spectrum at times $t= 50$, 75, 100
  calculated from the experimental data of higher resolution. 
  The shift of the curves for different times is apparent and 
  characteristic of an {\em intrinsic} anomalous scaling.  
  The straight line has slope -2.58 and is in agreement with a
  the power decay $k^{-(2\chi_{loc} + 1)}$ with $\chi_{loc} = 0.79$.} 
\end{figure}

\begin{figure}
\centerline{
\epsfxsize=5cm
\epsfbox{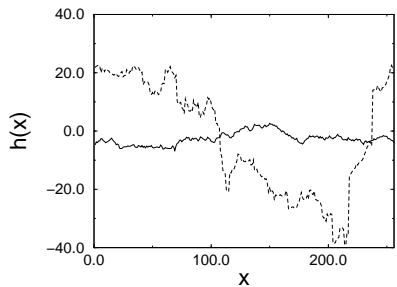}}
\caption{
Example of
two fractal curves with the same local roughness exponent but
different global one.
One interface (solid) has $\chi = \chi_{loc}=1/2$ and is
a realization of the simple $\partial_t h = \partial^2_x h + \xi$
equation, where $\xi$ is a Gaussian white noise, and it is thus a
true self-affine interface.
The other curve (dashed) has $\chi = 3/4$ and $\chi_{loc} = 1/2$ and
is a typical front of the random diffusion growth
process [17,19], which is well known to exhibit anomalous
roughening.
}
\end{figure}

\end{multicols}
\end{document}